# Spatial Data Infrastructures

**Yingjie Hu**, Department of Geography, University of Tennessee, Knoxville, TN 37996
**Wenwen Li**, School of Geographical Sciences and Urban Planning, Arizona State University, Tempe, AZ 85287

**Abstract:** Spatial data infrastructure (SDI) is the infrastructure that facilitates the discovery, access, management, distribution, reuse, and preservation of digital geospatial resources. These resources may include maps, data, geospatial services, and tools. As cyberinfrastructures, SDIs are similar to other infrastructures, such as water supplies and transportation networks, since they play fundamental roles in many aspects of the society. These roles have become even more significant in today's big data age, when a large volume of geospatial data and Web services are available. From a technological perspective, SDIs mainly consist of data, hardware, and software. However, a truly functional SDI also needs the efforts of people, supports from organizations, government policies, data and software standards, and many others. In this chapter, we will present the concepts and values of SDIs, as well as a brief history of SDI development in the U.S. We will also discuss the components of a typical SDI, and will specifically focus on three key components: geoportals, metadata, and search functions. Examples of the existing SDI implementations will also be discussed.

**Definitions**
**1. Spatial data infrastructure**: The technology, policies, standards, and human resources necessary to acquire, process, store, distribute, and improve utilization of geospatial data, services, and other digital resources.
**2. Geoportal**: A gateway website through which people can search, discover, access, and visualize the geospatial resources within a SDI.
**3. Metadata**: Documentation about who, when, how, what, why, and many other facets of the data and the data production process. Metadata can be used for describing not only data, but also tools, services, and other geospatial resources.
**4. Data standard**: A commonly agreed specification on how data should be recorded and described.
**5. Geospatial interoperability**: The ability of different geographic information systems to share, exchange, and operate (heterogenous) geospatial data and functions.
**6. Web service**: A Web application that provides standardized application programming interfaces to allow remote access to data and functions over the Internet.

___________________________

# 1. Spatial Data Infrastructure and its Values

The emergence of spatial data infrastructures (SDIs) is closely associated with the efforts of collecting and producing geospatial data, as well as the advancement of surveying and computer technologies. In the past decades, a large amount of geospatial data, such as remote sensing images and GPS locations, have been collected by government agencies such as the U.S. Geological Survey (USGS) and the National Oceanic and Atmospheric Administration (NOAA). Meanwhile, the fast development of geographic information systems facilitates the derivation of various data products from the collected data, such as topographic maps, land cover data, transportation networks, and hydrographic features. As location-based services are becoming increasingly popular, vast amounts of volunteered geographic information (VGI) (Goodchild 2007) has also been contributed by the general public through smart mobile devices and social media platforms. In addition, the componentization of GIS brings geospatial services that provide data processing and spatial analysis functions in the general Web environment. The large number of geospatial data, services, maps, and others, however, do not ease the use of these geospatial resources. On one hand, it is challenging to find and access these digital resources which are widely distributed at different government agencies and websites (Li, Wang and Bhatia 2016). On the other hand, a lot of data redundancies exist, and money and human resources were wasted in duplicated data collection and maintenance efforts (Rajabifard and Williamson 2001, Maguire and Longley 2005).

These problems were recognized by governments of the countries around the world, and many spatial data infrastructures were constructed since 1990s (Masser 1999). In the U.S., a national spatial data infrastructure (NSDI) initiative was started in 1993 to provide standardized access to geographic information resources (National Research Council 1993). An official definition of NSDI, according to the Executive Order 12906, is "the technology, policies, standards, and human resources necessary to acquire, process, store, distribute, and improve utilization of geospatial data." The Federal Geographic Data Committee (FGDC) is charged with coordinating the efforts to develop the NSDI in the U.S. The naming also indicates that SDI is recognized as an infrastructure similar to other types of infrastructures, such as electricity grids and water supplies, and that it plays fundamental roles in the socioeconomic and environmental developments of a country. Three parallel fronts were developed in the NSDI program: 1) a set of data standards for formalizing data and metadata; 2) a clearinghouse network providing data storage and online access; and 3) a set of framework data for the entire country, such as administrative boundaries (Longley et al. 2001, Maguire and Longley 2005).

Spatial data infrastructure presents a solution to the problems of resource discovery and data redundancy. It provides a unified platform where people can go and search geospatial data, maps, services, and other digital resources. As multiple government agencies are sharing their data on one platform, SDI reduces data redundancy and the extra efforts in collecting duplicated geospatial data. From a cost/benefit perspective, SDI allows geospatial data to be collected once and reused multiple times in different applications. More generally, SDI can be considered as an important element in the e-government (Georgiadou, Rodriguez-Pabón and Lance 2006) and open-government movement to increase the transparency of governmental activities and to enhance public participation. Better access to geospatial data also stimulates the growth of new businesses which may not be possible otherwise (Ralston 2004).

## 2. Key Components of a SDI

A SDI consists of many components. In addition to the digital geospatial resources, a SDI also needs hardware, software, people, organizations, standards, policies, and many others to function properly. Constructing a SDI also needs effective communications between communities, and negotiations among organizations and even countries to reach agreements. While a SDI has many components, this chapter will particularly focus on geoportals, metadata, and search functions, which are three key components of a typical SDI.

### 2.1 Geoportals

Geoportals are Web gateways that provide one-stop access to geospatial resources (Tait 2005). Geoportals are probably the most visible part of SDIs, since they are the main interfaces through which people can search and find geospatial resources. Figure 1 shows the user interface of Data.gov, a geoportal for the U.S. government open data. Figure 1(a) is the initial interface of the geoportal which contains a search bar that allows users to type in some keywords and find relevant resources. Figure 1(b) shows a list of the results when the keyword "watershed" is searched.

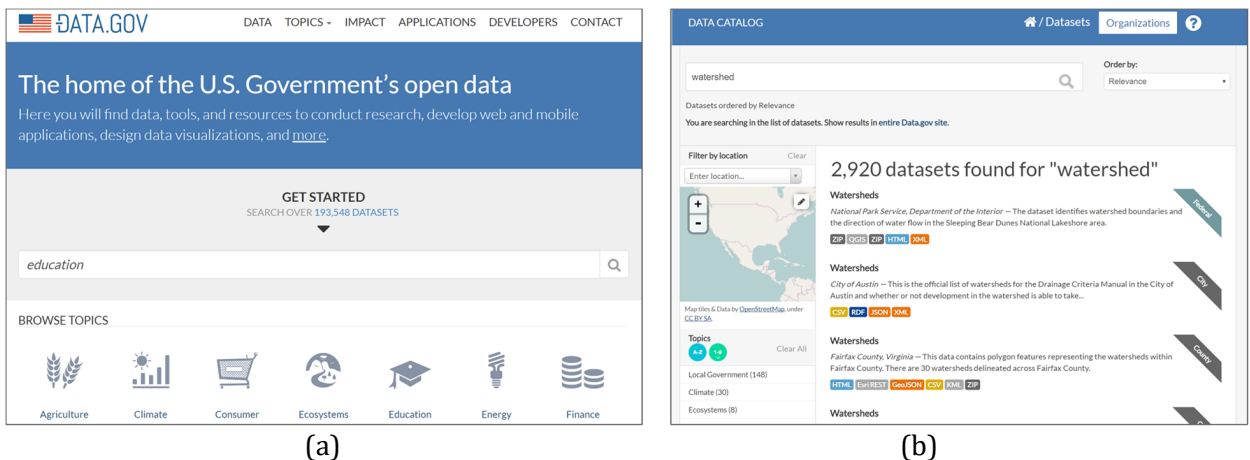

Figure 1. Two screenshots of the U.S. government open data geoportal Data.gov.

Geoportals are typically developed using Web-based technologies and off-the-shelf GIS software packages. A database management system (DBMS) is used to store and manage the metadata of the geospatial resources contained in the SDI. A Web interface, which often contains a map, enables end users to interact with the system and to conduct searches (Figure 2). When a search is performed, a HTTP (Hypertext Transmission Protocol) request will be sent to the Web server which hosts the geoportal. After querying the metadata stored in the database, the geoportal will then send back the result to the client through a HTTP response. Geoportals are typically designed to be used by both GIS professionals and the general public.

One important function of geoportals is helping users discover the existing geospatial resources. This resource discovery process often follows the publish-find-bind pattern (Rose 2004, Maguire and Longley 2005), in which: 1) providers publish the metadata of their data and services to a geoportal; 2) users perform a search on the geoportal and potentially find the data; 3) users consume the data and services from the providers. Figure 2 illustrates these three steps.

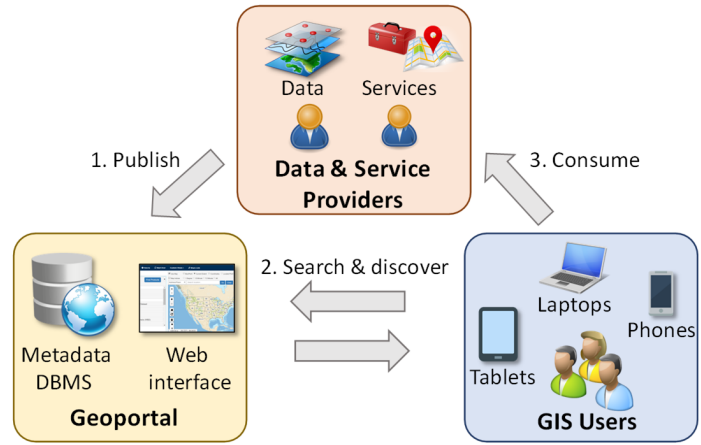

Figure 2. Some key components of a SDI and the publish-find-bind pattern.

The resource discovery process heavily relies on the quality of the metadata and the performance of the search function. In the following, we will discuss these two components.

## 2.2 Metadata

Metadata provide documentations on the content and the production process of geospatial resources. Metadata are often called the data about data, and include information such as titles, descriptions, data categories, the locations and time of the data collection, the data collectors, the used coordinate systems and map projections, and the data cleaning and processing procedures. Metadata can also be used for describing geospatial services by providing information about the data and functions offered by the services, the input and output, the developers, the development time, and others. In short, metadata are about all aspects of digital geospatial resources. Figure 3 shows the metadata fragment of a geospatial service provided by the NOAA, which describes the capabilities (e.g., getting a map based on the served geospatial data) and the data layers provided by this service.

```xml
<Capability>
  <Request>
    <GetCapabilities>...</GetCapabilities>
    <GetMap>...</GetMap>
    <GetFeatureInfo>...</GetFeatureInfo>
    <esri_wms:GetStyles>...</esri_wms:GetStyles>
  </Request>
  <Exception>...</Exception>
  <Layer>
    <Title>...</Title>
    <CRS>CRS:84</CRS>
    <CRS>EPSG:4326</CRS>
    <CRS>EPSG:3857</CRS>
    <!-- alias 3857 -->
    <CRS>EPSG:102100</CRS>
    <EX_GeographicBoundingBox>
       <westBoundLongitude>-179.999996</westBoundLongitude>
       <eastBoundLongitude>179.999996</eastBoundLongitude>
       <southBoundLatitude>-89.000000</southBoundLatitude>
       <northBoundLatitude>89.000000</northBoundLatitude>
    </EX_GeographicBoundingBox>
    <BoundingBox CRS="CRS:84" minx="-179.999996" miny="-89.000000" maxx="179.999996" maxy="89.000000"/>
    <BoundingBox CRS="EPSG:4326" minx="-89.000000" miny="-179.999996" maxx="89.000000" maxy="179.999996"/>
    <BoundingBox CRS="EPSG:3857" minx="-20037507.842788" miny="-30240971.458386" maxx="20037507.842788" maxy="30240971.458386"/>
```

Figure 3. The metadata fragment (in the format of XML) of a geospatial service provided by the NOAA.

Metadata are fundamentally important for SDIs. When data and services leave their original data production context and are integrated into a SDI, metadata provide the primary information based on which GIS users can understand and use digital geospatial resources. Without metadata or with only poorly constructed metadata, it is very difficult, if not impossible, for data and services to be reused. The quality of metadata also determines the result of resource discovery. Many geoportals rank the relevance of geospatial resources to user queries based on the information contained in their metadata. Complete and accurate metadata allow geoportals to find and rank geospatial resources based on locations, time, thematic attributes, data types, published years, data collectors, and many other conditions explicitly or implicitly specified in the user queries. To ensure the quality of metadata, standards are established to define the necessary elements that should be included in metadata. In the U.S., the FGDC is responsible for coordinating the development of metadata standards, and its Content Standard for Digital Geospatial Metadata (CSDGM) has been used by many U.S. government agencies to formalize metadata. Since 2010, FGDC has endorsed a series of international metadata standards (e.g., the ISO 191** standards) to promote Global Spatial Data Infrastructure (GSDI).

## 2.3 Search Function

Search functions are the major means through which users discover geospatial resources in a SDI. Without an effective search function, relevant geospatial resources in an SDI can hardly be found by the users. Two types of search functions are often adopted in geoportals: text-based search and map-based search. Text-based search is similar to Web search engines, in which a user types in some keywords and receives the results based on the matched text. Map-based search allows users to find geospatial resources by interacting with a map, and a user can pan, zoom in and out, and draw polygons to specify their areas of interest. Using either of these two methods alone has some limitations. The text-based search enables general users, especially the users who are unfamiliar with a GIS, to find geospatial resources in a way similar to how they would use a general search engine, such as Google. However, it can be challenging to identify the suitable keywords, or the keywords may not accurately describe the geographic areas that the user is interested in. The map-based search, on the other hand, provides convenience for the users who are already familiar with a map interface, and allows users to specify geographic locations in a more accurate manner (e.g., by drawing polygons). However, not everyone feels comfortable of using a map-based interface. Geoportals often offer both search functions to complement the two and accommodate the needs of different users

The search functions in SDIs are being improved by researchers. One of these improvements lies in text-based search: there is a transition from keyword-based search to semantic search. Keyword-based search examines the matching between the keywords input by the users and the textual descriptions of the geospatial resources. Thus, if a user types in "road", the search function will not be able to find the resources labeled as "street". Semantic search aims to match digital geospatial resources to user queries based on the meaning (semantics) of the queries, and therefore can identify relevant data and services even though they are not labeled with the exactly same words. Early attempts of semantic search include the use of ontology-based query expansion (Lutz and Klien 2006), rule-based semantic reasoning (Li, Zhou and Wu 2016), and faceted search, e.g., Apache Solr (Hostetter 2006). Despite the existing research efforts, additional work is still necessary to integrate semantic search into current SDIs. Figure 4 shows examples of searching "earthquake" (Figure 4(a)) and "natural disasters" (Figure 4(b)) respectively on Data.gov. While earthquake is generally considered as a type of natural disaster, the search of "earthquake", however, returned more matching records than the other search, suggesting that the search function behind is more likely to be based on a simple keyword matching. In addition to text-based search, map-based search is also supported by Data.gov. Figure 4(c) shows a spatial filtering effect by narrowing down the region of interest to the contiguous United States using the map interface.

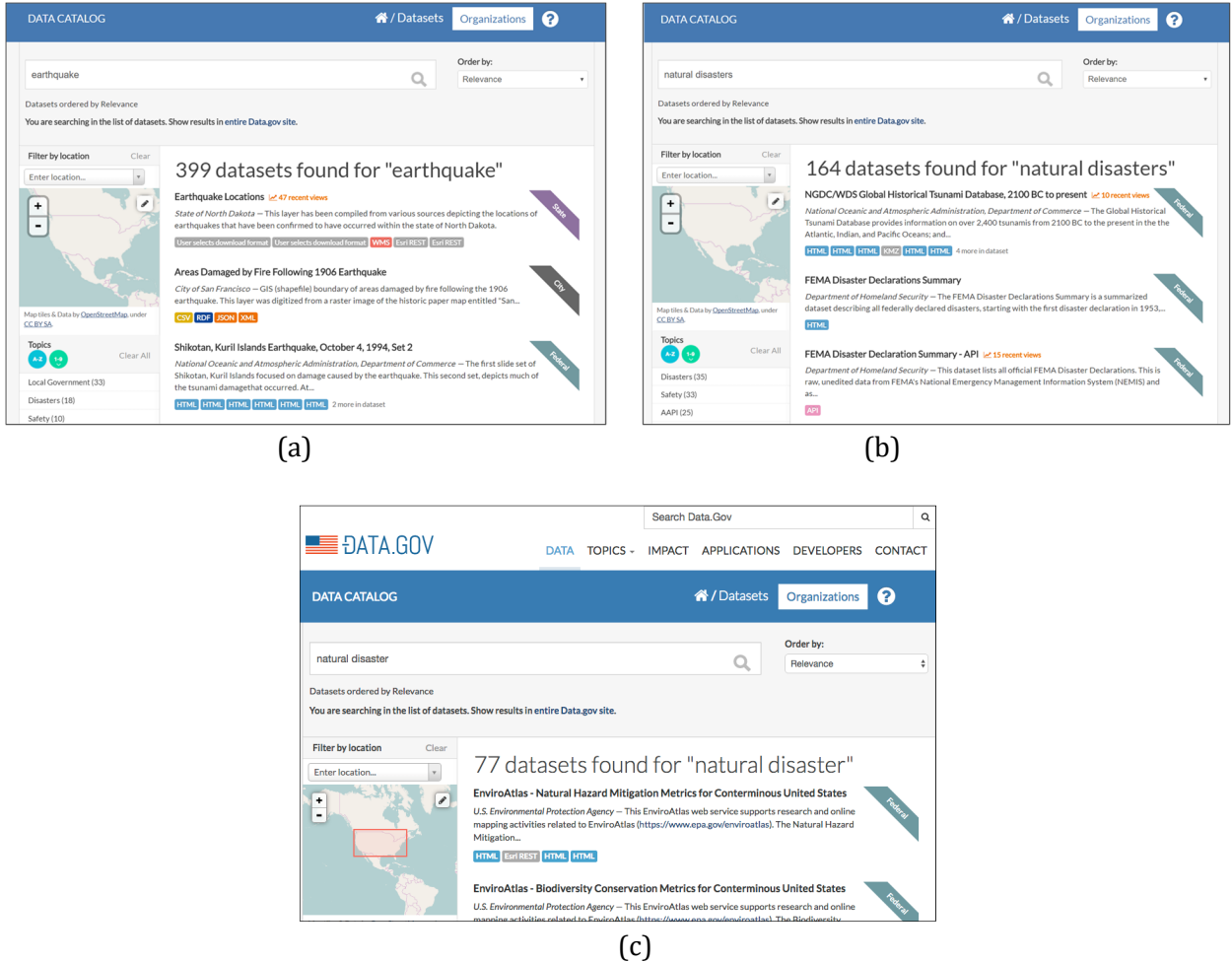

Figure 4. Search examples based on Data.gov.

**3. Examples of SDIs**

There are many spatial data infrastructures established at different geographic levels. At the global level, there is Global Earth Observation System of Systems which combines the efforts from more than 70 countries to share environmental data. At the continental level, there is Infrastructure for Spatial Information in the European Community (INSPIRE) which enables the sharing of geospatial information among public organizations across Europe. At the national level, there is Data.gov which provides access to a large number of United States government open datasets. The National Map (TNM) is a USGS SDI project which supports easy access and downloading of topographic information about elevation, geographic names, hydrology, boundaries, transportation, and so forth. There are also national SDIs in Australia, China, Japan, Malaysia, Netherlands, Portugal, and other countries. At the state level, there is Tennessee GIS portal which offers the geospatial datasets, services, and Web applications related to the State of Tennessee. At the city level, there is New York

City Open Data portal . There are also spatial data infrastructures dedicated to specific domains, such as disaster response, public health, and climate change.

**4. Current and Possible Future Developments of SDIs**
Spatial data infrastructures heavily rely on computer and information technologies, and are continuously evolving with the technological advancements. Techniques commonly used in today's SDIs, such as Asynchronous JavaScript and XML (AJAX) which enables asynchronous processing to speed up search performance as well as user experience, are not available for the first generation of SDIs in 1990s. Similarly, we may see the emergence of new technologies that can improve SDIs in various aspects, and some of these technologies are already being tested in research labs. For example, graph databases, such as Neo4j, may be employed to augment the existing relational databases that have been widely used in SDIs to store metadata. Semantic Web (or the third generation of the Web) and Linked Data technologies may be leveraged to turn SDIs into local Semantic Webs (Athanasis et al. 2009). The concept of knowledge graph (Singhal 2012) may be applied and integrated to SDIs to construct knowledge bases that link the geospatial entities and concepts contained in the data. Machine learning and data mining methods, such as latent semantic analysis and labeled latent Dirichlet allocation may help automatically infer and generate high-quality metadata from the content of maps and services (Li, Bhatia and Cao 2015, Hu et al. 2015). While there can be lots of possible technological improvements, future SDIs also need the critical support from people, organizations, and governments. With these important components, a future SDI will contribute more to the development of the society.

**Additional Resources**
1. The Federal Geographic Data Committee: http://www.fgdc.gov
2. Open government movement: https://obamawhitehouse.archives.gov/open
3. Global Earth Observation System of Systems: http://www.geoportal.org
4. Infrastructure for Spatial Information in the European Community: http://inspire.ec.europa.eu
5. Data.gov: https://www.data.gov
6. The National Map: https://nationalmap.gov
7. Tennessee GIS portal: http://tnmap.tn.gov
8. New York City Open Data portal: https://nycopendata.socrata.com

**Learning Objectives**
1. Define spatial data infrastructure.
2. Describe the major components of a typical SDI.
3. Describe the main functions of geoportals.
4. Define metadata, and describe the types of information that may be included in metadata.
5. Differentiate text-based search and map-based search.
6. List some of the widely-recognized SDIs.
7. Describe the role of standards in ensuring the quality of metadata.

**Instructional Questions**
1. What are the societal benefits of spatial data infrastructures? In which aspects can SDIs improve the sharing and reusing of geospatial data and services?
2. What are the advantages and disadvantages of text-based and map-based searches? What are the differences between keyword-based and semantic searches?
3. How can metadata benefit the understanding and use of geospatial data and services? How can metadata facilitate the search and discovery of digital geospatial resources?